\documentclass[pre,aps,amsmath,amsfonts,amssymb,twocolumn,showpacs]{revtex4}
\usepackage{graphicx}
\usepackage{bm}
\usepackage{dcolumn}
\def\brho{{\bm \rho}}
\def\bM{{\bm{M}}}
\begin{document}
\title{Cluster mean-field approximations with the coherent-anomaly-method 
analysis for the driven pair contact process with diffusion}
\author{Su-Chan Park}
\affiliation{School of Physics, Korea Institute for Advanced Study, 
Seoul 130-722, Korea}
\author{Hyunggyu Park}
\affiliation{School of Physics, Korea Institute for Advanced Study, 
Seoul 130-722, Korea}
\date{\today}
\begin{abstract}
The cluster mean-field approximations 
are performed, up to 13 cluster sizes, to study
the critical behavior of the driven pair contact process with diffusion (DPCPD)
and its precedent, the PCPD in one dimension. Critical points 
are estimated by extrapolating our data
to the infinite cluster size limit, which are in good accordance 
with recent simulation 
results. Within the cluster mean-field approximation scheme, 
the PCPD and the DPCPD 
share the same mean-field critical behavior. The application of 
the coherent anomaly method,
however, shows that the two models develop different coherent anomalies, 
which lead to different true critical scaling. The values of 
the critical exponents for the particle density, the pair density, 
the correlation length, and the relaxation time
are fairly well estimated for the DPCPD. These results support and complement 
our recent simulation results for the DPCPD. 
\end{abstract}
\pacs{05.70.Ln, 02.60.-x, 64.60.Ht}
\maketitle

The absorbing phase transition (APT) has been studied extensively 
to understand many-body cooperative phenomena in nonequilibrium systems 
\cite{MD99}. Up to now, two universality classes have been firmly 
established: directed percolation (DP) and parity conservation (PC) 
universality classes \cite{H00}. A few other candidates for different universality
classes have been reported in recent literatures. One is the 
DP system coupled with a static conserved field \cite{RPV00}. Although
the reported values for the critical indices are rather 
scattered \cite{KC1,PPun,privateC}, 
it is widely believed that these systems form 
a universality class, different from the DP 
and the PC class. Another candidate is 
the pair contact process with diffusion (PCPD)
that has as yet defied any consensus on the universality issue. 
Various scenarios have been proposed, including a new single universality 
class \cite{Hin010,KC03},
a marginally perturbed DP process with continuously 
varying exponents \cite{Noh04}, and
a DP process with a huge crossover time \cite{Hin030,BC03}, 
which are summarized in a recent review \cite{HH04}.

Recently, we studied the driven PCPD (DPCPD) which is a variant of the PCPD 
by introducing biased diffusion \cite{PP04}. 
It is shown that the driving is relevant 
and the DPCPD exhibits a ``mean-field-like" critical 
behavior even in one dimension.
Since the DP class is insensitive to the driving, the DP scenario 
with a huge crossover time should be eliminated. 
There was a recent attempt to understand the PCPD using the renormalization
group (RG) analysis on a single-species Bosonic 
action derived from the microscopic
master equation. However, it turned out to be improper 
to describe the critical behavior of the
PCPD \cite{JvWOT04}.  In our previous work \cite{PP04}, 
we pointed out a possible
reason for this failure and suggested that the PCPD may be described properly 
by a field theory with two independent fields. 
Still, the search for the coarse-grained action adequate for 
the PCPD remains a challenge. 

Besides the RG technique on the proper action \cite{RGall}, 
there are a few other efficient methods 
to investigate the absorbing critical phenomena. Numerical simulations along
with a finite-size-scaling analysis \cite{Aukrust} and direct 
integrations of corresponding Langevin
equations \cite{DCA} are two typical examples. 
Another frequently used method is the
cluster mean-field (CMF) approximation \cite{bAK} 
followed by the coherent-anomaly 
method (CAM) analysis \cite{CAM}. This method is known to be effective 
to obtain a
quantitative  phase diagram and sometimes even explore a true critical 
scaling behavior \cite{CAMapp}.
However, the accurate measurement of critical 
indices is only limited to rather simple DP systems.
More complex critical behaviors like 
in the PC and the PCPD classes could not have been probed 
with a reasonable accuracy as yet by the CAM analysis \cite{CAMsol,Szol2}. 

In this paper, motivated by our recent results that the DPCPD exhibits a 
distinct critical behavior from the PCPD
and also a mean-field-like behavior even 
in one dimension \cite{PP04},  we develop the CMF 
approximations for the DPCPD and the PCPD, expecting 
that the CAM analysis would produce
a reasonable estimate for the mean-field-like critical 
indices of the DPCPD. Also, direct comparison of
the CMF data for two models may provide an independent 
support for different scaling behaviors.

We set up dynamic CMF equations up to $n=13$ cluster size. The steady-state
solutions are obtained within machine accuracy  using {\sc mathematica}. 
Dynamic information is also extracted from the 
smallest eigenvalue of the ``linearized''
transition matrix. Subsequently, through the 
CAM analysis, we estimate the values of the critical exponents
for the particle density, the pair density, the correlation 
length, and the relaxation time.

The model is defined on a one-dimensional lattice of $L$ sites 
with periodic boundary conditions. 
At each site, there is at most one particle and no 
multiple occupancy is allowed. Hence the configuration is specified 
by the occupation number which is either 1 or 0 at every site. 
Each particle hops to the right (left) with transition rate $D_R$ ($D_L$).
The total number of particles in the system varies by branching 
and annihilating events mediated by a particle pair 
($2A\rightarrow 3A$ and $2A\rightarrow \emptyset$).
The transition rate is $p$ ($1-p$) 
for the annihilating (branching) event with $0\le p\le 1$. 
These three dynamics can be described by the master equation which 
takes the form
\begin{equation}
\frac{\partial}{\partial t} | P;t\rangle = - \hat H | P ; t\rangle,
\label{Eq:master}
\end{equation}
where $| P;t\rangle$ is the state vector at time $t$ and 
the ``Hamiltonian'' is written as $\hat H = \sum_{i=1}^L \hat H_i$ with
\begin{equation}
\begin{aligned}
\hat H_i &=  D_R ( \hat \rho_i \hat v_{i+1}-\hat a_i \hat a_{i+1}^\dag  )
+ D_L ( \hat v_i \hat \rho_{i+1}  -\hat a_i^\dag \hat a_{i+1} )\\
&- p ( \hat a_i \hat a_{i+1} - \hat \rho_i \hat \rho_{i+1} ) \\
&- \frac{1-p}{2} ( \hat a_{i-1}^\dag + \hat a_{i+2}^\dag 
 - \hat v_{i-1} -\hat v_{i+2} ) \hat \rho_i \hat \rho_{i+1},
\label{Eq:Hamiltonian}
\end{aligned}
\end{equation}
where $\hat a_i (\hat a_i^\dag) $ is the annihilation (creation) operator
of hard core particles, satisfying 
$\{ \hat a_i, \hat a_i^\dag \}=1$ 
and $ [\hat a_i, \hat a_j] = [\hat a_i, \hat a_j^\dag]=0$ for $i\neq j$,
$\hat \rho_i =
\hat a_i^\dag \hat a_i$ is the number operator, 
and  $\hat v_i = 1 - \hat \rho_i$.

Three different cases arise depending on the values of $D_R$ and $D_L$.
The case of $D_R = D_L = 0$ represents the pair contact process (PCP) which
has infinitely many absorbing states and  is known to belong to the
DP class at least for static situations \cite{J93}. 
Since the cluster approximations along with the CAM analysis  have been
already performed for the PCP by several authors previously \cite{PCP}, 
we skip the analysis of the PCP here. The PCPD corresponds to
$D_R  = D_L \neq 0$ and the DPCPD corresponds to $D_R \neq D_L$.
In what follows, we set $D_R + D_L = 1$ for convenience 
and $D_R$ is chosen  to be 1/2 (1) for the PCPD (DPCPD).

We consider an $n$-site probability function $P_n({\brho};t)$.
It is defined as the probability at time $t$ to find an $n$-site cluster of 
the configurational state ${\brho}=(\rho_1, \rho_2, \ldots, \rho_n)$,
where an occupational state $\rho_i$ at site $i$ takes either 0 or 1.
Tracing out Eq.~(\ref{Eq:master}) over occupational states outside the cluster 
($\{\rho_i$\} with $i\le0$ or $i\ge n+1$), 
one may find a formal exact expression
\begin{equation}
\frac{d P_n(\brho;t)}{dt}  = \tilde F_\brho( P_n, P_{n+1}, P_{n+2}),
\label{Eq:exact_rate}
\end{equation}
where the function $\tilde F_\brho$ involves the sets 
of $n$-, $(n+1)$-, and $(n+2)$-site probability. Notice that 
$P_{n+1}$ and $P_{n+2}$ terms show up due to the boundary dynamics
of the $n$-site cluster. 

As the infinite hierarchy appearing in 
Eq.~(\ref{Eq:exact_rate}) is the major obstacle towards analytic
treatment, we need an approximation scheme to truncate the hierarchy 
at finite $n$. In this paper, we take the 
so-called $(n+1,n)$ approximations \cite{bAK},
where $P_{n+2}$ and $P_{n+1}$ are expressed in terms of products of $P_{n}$'s. 
Then, the rate equations for the $n$-site cluster probability function become
\begin{equation}
\frac{d P_n (\brho;t) }{dt} = F_{\brho} ( \{P_n\}),
\label{Eq:rate}
\end{equation}
where $P_n$ is now the approximate (mean-field) probability function.

The stationary probability distribution function $P_n^s(\brho)$
can be obtained by solving the set of coupled equations $F_{\brho} = 0$. 
For given $n$, the number of equations and 
the number of variables are both $2^n$, but
not all are independent. The translational invariance and the 
normalization condition
guarantee that all $P_n(\brho)$ with $\rho_1=0$ can be expressed in a linear
combination of $P_n(\brho)$'s with $\rho_1=1$. 
For example, in case of $n=4$, 
one can easily show that $P_4(0011) = P_3(011) - P_4(1011)
=P_2(11)-P_3(111) - P_4 (1011)=P_3(110)-P_4(1011)
=P_4(1101)+P_4(1100) - P_4(1011)$. 
The absorbing (vacuum) probability $P_4(0000)$ can
be determined by the normalization condition. 
Hence the DPCPD case has $2^{n-1}$ independent variables.
In case of the PCPD, the left-right symmetry further reduces the number of
independent variables, for example, $P_4(1101) = P_4(1011)$ and so on.

We use {\sc mathematica} to find the stationary solutions $P_n^s(\brho)$ up to $n=13$
with machine accuracy ($10^{-20}$) for given values of parameters, 
$p$ and $D_R$  \cite{mathematica}. With $P_n^s$,
we calculate the particle density
$\rho_s$ and the pair density $\rho_p$ in the steady state as
\begin{eqnarray}
\rho_s &=& P_1^s(1) = {\rm Tr}_{\brho} ~P_n^s(\brho) \delta_{\rho_k,1}, \nonumber\\
\rho_p &=& P_2^s(11) = {\rm Tr}_{\brho} ~P_n^s(\brho) \delta_{\rho_k,1} \delta_{\rho_{k+1},1},
\end{eqnarray}
where $k$ (and $k+1$) denotes an arbitrary site inside the cluster. 
At a fixed value of $D_R$, the order parameters, $\rho_s$ 
and $\rho_p$, simultaneously vanish
for large $p$ (pair annihilation rate) 
and the system exhibits an absorbing phase transition
into vacuum at $p=p_c^n$.  

\begin{figure}[t]
\includegraphics[width=0.48\textwidth]{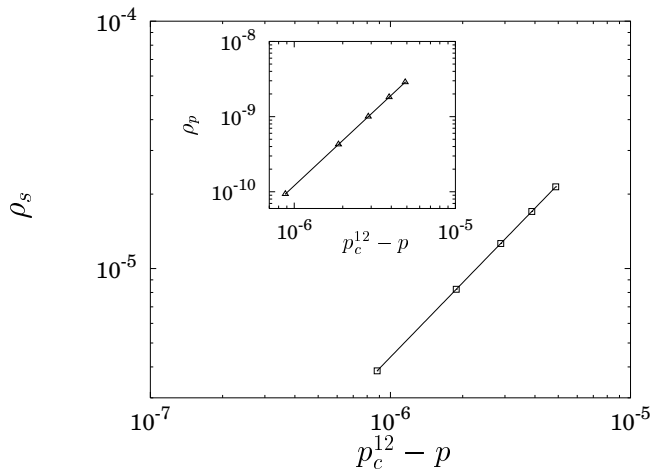}
\caption{\label{Fig:example} Log-log plot of $\rho_s$ vs $p_c^{12} -p$
obtained from the 12-cluster CMF approximation for the DPCPD. The slope 
of the straight line is 1. In the inset, $\rho_p$'s are plotted 
against $p_c^{12} -p$. The slope of the straight line is 2.}
\end{figure}

Near the transition point $p_c^n$, the order parameters scale as
\begin{equation}
\rho_s \simeq A_n ( p_c^n - p )^{\beta_1^\text{MF}},\quad 
\rho_p \simeq B_n (p_c^n - p)^{\beta_2^\text{MF}},
\end{equation}
where we find the mean-field values for the order parameter exponents:
$\beta_1^\text{MF}=1$ and  $\beta_2^\text{MF}=2$.
Figure \ref{Fig:example} shows the $n=12$ cluster approximation results
for the DPCPD.  We estimate the critical point $p_c^n$ 
and the critical amplitudes $A_n$ and $B_n$
by fitting five data near the transition ($|p-p_c^n|\le 5\times 10^{-6}$),
linearly for $\rho_s$ and quadratically for $\rho_p$. 
Our results are tabulated in Table \ref{Table:CAM_PCPD} for the PCPD model
and in Table \ref{Table:CAM_DPCPD} for the DPCPD model.
Notice that the relative errors  for $p_c^n$ 
are extremely small ($\sim10^{-9}$), but
the amplitudes $A_n$ and $B_n$ still 
have a sizable relative error  ($\sim 10^{-4 }$). 

It is interesting to note that, for $n\le 3$, the diffusion bias does not enter 
the CMF rate equations at all. The functional form of $F_\brho (\{ P_n\})$
in Eq.~(\ref{Eq:rate}) is identical for the PCPD and the DPCPD. 
The left-right symmetry among $P_n(\brho)$'s is automatically enforced
due to the translational invariance, regardless of the details of the dynamics. 
For example, $P_3(110) = P_2(11) - P_3(111) = P_3(011)$ and so on. However, 
for $n\ge 4$, the translational invariance does 
not guarantee the left-right symmetry,
which may be broken by the dynamics with a broken left-right symmetry.

\begin{table}[t]
\caption{\label{Table:CAM_PCPD} Cluster approximation results for
the PCPD model. The errors are in the last digits.}
\begin{ruledtabular}
\begin{tabular}{ccccc}
$n$&$p_c^n$&$A_n$&$B_n$&$C_n$\\
\hline
4 &0.209~692~7263&4.473&51.855&17.59\\
5 &0.194~357~9912&4.720&72.928&19.15\\
6 &0.184~167~8676&4.859&93.789&19.93\\
7 &0.177~119~7696&4.963&116.26&20.66\\
8 &0.171~815~3824&5.039&139.91&21.22\\
9 &0.167~700~6591&5.100&165.04&21.72\\
10&0.164~396~9333&5.151&191.63&22.17\\
11&0.161~685~1815&5.194&219.71&22.58\\
12&0.159~416~2244&5.232&249.28&22.96\\
13&0.157~488~7140&5.265&280.35&     \\
\end{tabular}
\end{ruledtabular}
\end{table}
\begin{table}[b]
\caption{\label{Table:CAM_DPCPD} Cluster approximation results for
the DPCPD model. The errors are in the last digits.}
\begin{ruledtabular}
\begin{tabular}{ccccc}
$n$&$p_c^n$&$A_n$&$B_n$&$C_n$\\
\hline
4 &0.216~140~3513&4.254&44.37&16.69\\
5 &0.202~800~9465&4.356&56.08&17.31\\
6 &0.194~381~7410&4.405&66.48&17.55\\
7 &0.188~503~1907&4.423&76.39&17.74\\
8 &0.184~102~4774&4.424&85.76&17.83\\
9 &0.180~689~8311&4.420&94.89&17.90\\
10&0.177~954~3360&4.410&103.7&17.94\\
11&0.175~711~7674&4.397&112.3&17.97\\
12&0.173~837~8803&4.383&120.8&17.99\\
13&0.172~247~8976&4.368&128.9&     \\
\end{tabular}
\end{ruledtabular}
\end{table}

In nonequilibrium systems, dynamic relaxation behavior provides one of the key
pieces of information on the system. Off criticality, 
the order parameters are expected to approach 
their stationary values exponentially with a 
characteristic relaxation time $\tau$. 
At criticality, $\tau$ diverges and the order parameters decay algebraically.
One may roughly estimate $\tau$ by numerically integrating the rate equations 
(\ref{Eq:rate}) and fitting time-dependent data into an exponential form. 
However, this method does not produce high-precision data. 
In this paper, we propose a different method to calculate $\tau$ with machine accuracy
in the CMF approximation scheme. 

Since the stationary solutions of Eq.~(\ref{Eq:rate}) 
were obtained  with machine accuracy,
we can linearize Eq.~(\ref{Eq:rate}) near the 
stationary solutions very accurately.
The linearized equation takes the form
\begin{equation}
\frac{d | P_n;t\rangle}{dt} =- \bM | P_n;t\rangle,
\end{equation}
where $| P_n;t\rangle$ is the ($n$-cluster) state vector with the components 
$P_n(\brho;t)$ and $\bM$ is a square matrix. It is trivial to show that 
the eigenvalues of $\bM$ are equal to the 
inverse of various characteristic time scales
of the dynamics. The most dominant slow mode is determined by the smallest
eigenvalue $\Lambda_s$, i.e., the relaxation time $\tau=\Lambda_s^{-1}$.

We analyze the linearized equation up to $n=12$.  Near criticality, we find 
\begin{equation}
\tau^{-1} \simeq C_n ( p_c^n - p )^{\nu_{\|}^\text{MF}},
\end{equation}
where we find again the mean-field value for the relaxation exponent,
$\nu_{\|}^\text{MF}=2$. We estimate the critical 
points $p_c^n$ independently, which
are found to be consistent with previous estimates from the density data in 
Tables \ref{Table:CAM_PCPD} and \ref{Table:CAM_DPCPD}, where we also tabulate
the estimated values for the amplitude $C_n$ for both the PCPD and the DPCPD.

\begin{figure}[t]
\includegraphics[width=0.45\textwidth]{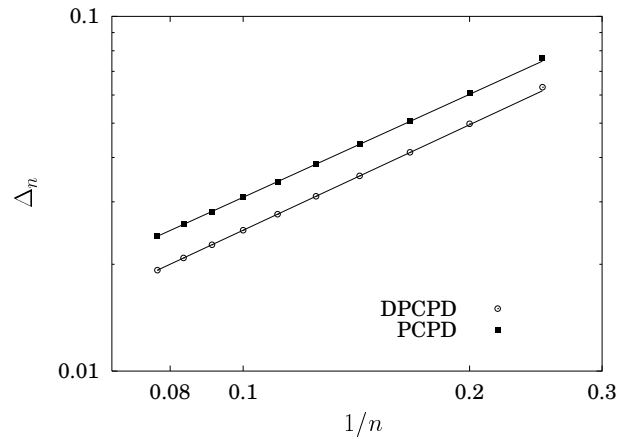}
\caption{\label{Fig:cam_nuperp} Log-log plots of $\Delta_n$ vs 
$1/n$ for the PCPD and the DPCPD.  In this figure, the value of $p_c$ 
for the PCPD (DPCPD) is set to be 0.1335 (0.153).}
\end{figure}

Now, we employ the coherent-anomaly method (CAM) introduced by Suzuki and
co-workers \cite{CAM} and estimate the values of the true critical exponents.
Following the CAM analysis, the $n$ dependence of the critical point $p_c^n$
is predicted in the large $n$ limit as
\begin{equation}
\Delta_n^{\nu_\perp} \sim n^{-1}, \label{Eq:nuperp}
\end{equation}
where $\Delta_n=p_c^n-p_c$ is the distance of $p_c^n$ from the 
true critical point $p_c=\lim_{n\rightarrow\infty} p_c^n$ and
$\nu_\perp$ is the true correlation length exponent.

We estimate $p_c$ by applying the Bulirsch and Stoer (or BST) 
algorithm \cite{HS88}
to the series of $\{p_c^n\}$ and find that $p_c=0.134(2)$ for the PCPD
and $p_c=0.154(3)$ for the DPCPD, which are in good agreement with
simulation results of 0.133~522(2) and 0.151~032(1) \cite{PP04}.
Alternatively, we estimate $p_c$ and $\nu_\perp$ simultaneously using
Eq.~(\ref{Eq:nuperp}).  In Fig. \ref{Fig:cam_nuperp}, we plot 
$\Delta_n$ vs $1/n$ in a log-log plot, varying $p_c$
to find the best power-law fit. For the PCPD, the choice of 
$p_c=0.1335$ yields the smallest fitting error with $\nu_\perp=1.04$,
where the data from $n=8$ to $13$ are used. 
For the DPCPD, the best choice is $p_c=0.153$ with $\nu_\perp=1.01$.
The relative error for $p_c$ is $\sim 2\%$, and  the error
for $\nu_\perp$ is $\sim 10\%$. The best simulation result of
$\nu_\perp=1.09(2)$ for the PCPD is within the errors, but the accurate 
measurement seems to be out of reach with data up to $n=13$. Our estimate
of $\nu_\perp=1.01$ for the DPCPD is in very good agreement with 
the expected mean-field value $\nu_\perp^\text{MF}=1$.

\begin{figure}[t]
\includegraphics[width=0.45\textwidth]{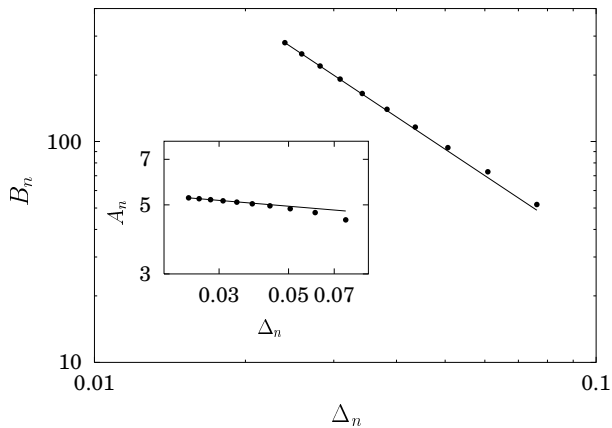}
\caption{\label{Fig:PCPD_beta} The CAM analysis for the order parameters 
for the PCPD model. 
The slope of the straight line is $-1.51$, which leads to $\beta_2\approx 0.49$.
In the inset, the slope of
the straight line is $-0.084$, which leads to $\beta_1\approx 0.92$.}
\end{figure}

The amplitudes $A_n$ and $B_n$ are expected to scale as
\begin{equation}
A_n \sim \Delta_n^{-({\beta_1^\text{MF} - \beta_1})}, \quad
B_n \sim \Delta_n^{-(\beta_2^{\text{MF}} - \beta_2)}, \label{Eq:beta}
\end{equation}
where $\beta_1$ and $\beta_2$ are the true critical exponents
for the order parameters.
In Fig. \ref{Fig:PCPD_beta}, 
the log-log plots of $A_n$ and $B_n$ vs $\Delta_n$ 
for the PCPD are presented. 
Here we use $p_c=0.133~522$ (the best estimate from Monte 
Carlo simulations) \cite{PP04}. The CAM analysis leads to $\beta_1\approx 0.92$
and $\beta_2\approx 0.49$ \cite{Szol3}, 
both of which are far from the simulation results
of $\beta_1\approx\beta_2\approx 0.36(2)$ \cite{KC03,Noh04}. 
In particular, there is a huge discrepancy between the estimated
values of $\beta_1$ and $\beta_2$ by the CAM analysis,
which warns us that the CAM estimates for the order parameter exponents
should be interpreted with great caution.
This huge discrepancy also implies that the cluster sizes 
up to $n=13$ are still too
small for the PCPD to reach the asymptotic regime where the system is 
dominated by long spatial correlations, induced by
the long-term memory mediated by solitary particles~\cite{Noh04}.

\begin{figure}[t]
\includegraphics[width=0.45\textwidth]{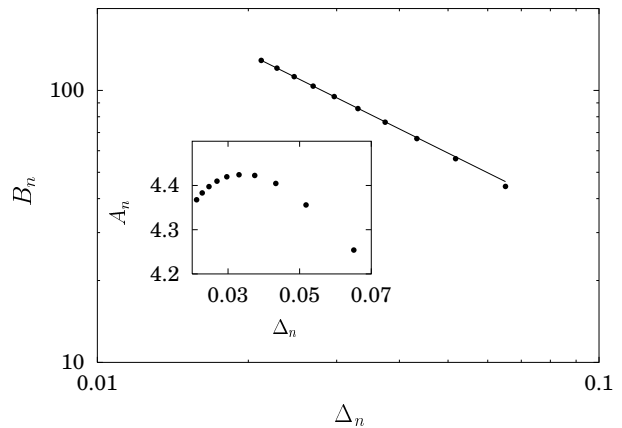}
\caption{\label{Fig:DPCPD_beta} 
The CAM analysis for the order parameters for the DPCPD model. 
The slope of the straight line is $-0.92$, which leads to $\beta_2\approx 1.08$.
In the inset, $A_n$ vs $\Delta_n$ is drawn without a log scale.
$A_n$ remains nearly constant which implies 
$\beta_1 \approx \beta_1^\text{MF}= 1$.}
\end{figure}
\begin{figure}[b]
\includegraphics[width=0.45\textwidth]{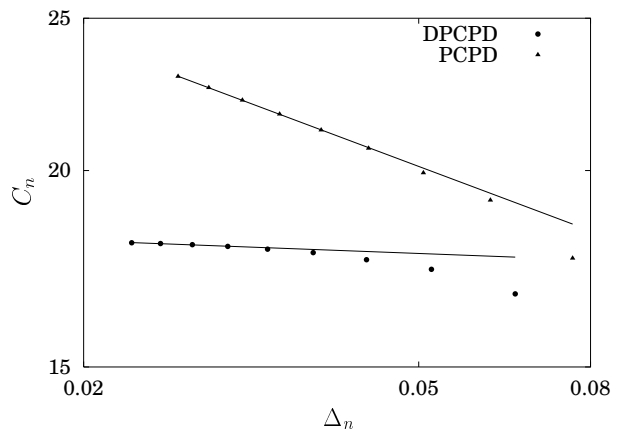}
\caption{\label{Fig:nupara} The CAM analysis for the relaxation time 
for the PCPD and the DPCPD.
The slope of the  straight line for the PCPD (DPCP) is $-0.2$ ($-0.02$)
which leads to $\nu_\| = 1.8~ (1.98)$.}
\end{figure}

On the other hand, the CAM analysis for the DPCPD looks consistent with 
the simulation results.
In Fig. \ref{Fig:DPCPD_beta}, we use $p_c = 0.151~032$ \cite{PP04}. 
First, $A_n$ seems not diverging as $\Delta_n\rightarrow 0$ and 
reaching a nonzero constant, which implies $\beta_1 = \beta_1^\text{MF}=1$. 
Second, $B_n$ behaves very differently from $A_n$ and diverges with the
exponent $\sim 0.92$, which implies that 
$\beta_2=\beta_2^\text{MF}-0.92\approx 1.08$. Numerical
simulation results \cite{PP04} are in complete agreement with our CAM results.
One should notice that $\beta_2$ does not assume the MF value, but seems to
be equal to $\beta_1$ except a probable multiplicative logarithmic correction
as found in the exponent $\beta/\nu_{||}$ by numerical simulations ~\cite{PP04}.
This mean-field-like behavior is expected for the 
two-dimensional PCPD~\cite{2dPCPD}, of which
the upper critical dimension is believed to be 2. Our CAM results independently 
support the conclusion drawn from our numerical simulations results \cite{PP04} 
that the DPCPD critical behavior is distinct from the PCPD behavior and
the upper critical dimension for the DPCPD is 1 rather than 2. 

Finally, we estimate $\nu_\|$ from the relation
\begin{equation}
C_n \sim \Delta_n^{-(\nu_\|^{\text{MF}} - \nu_\| )},
\end{equation}
where $\nu_\|$ is the true critical exponent for the relaxation time.
In Fig. \ref{Fig:nupara}, the log-log plots of $C_n$ 
vs $\Delta_n$ are shown.
We estimate that $\nu_\| \approx 1.8$ for the PCPD and 
$\nu_\| \approx 1.98$ for the DPCPD. Rather surprisingly, the PCPD result
is consistent with the simulation result of 
$\nu_{\|}=1.85(10)$ \cite{KC03,Noh04}.
For the DPCPD, the value of $\nu_{\|}$ is quite close to the mean-field value of
$\nu_{\|}^\text{MF}=2$, consistent with the simulation results.

In summary, we estimated the critical exponents for the PCPD and the DPCPD,
using CMF approximations along with the CAM analysis.
For the PCPD, the values of the order parameter 
exponents are poorly estimated, while
the estimates for the correlation and the relaxation exponents are consistent
with simulation results within error bars. 
In contrast, the CAM estimates for the
DPCPD are in excellent accord with simulation results, supporting our
conjecture that the upper critical dimension of the DPCPD is 1.


\begin{thebibliography}{99}
\bibitem{MD99} J. Marro and R. Dickman, {\it Nonequilibrium Phase Transitions
in Lattice Models} (Cambridge University Press, Cambridge, England,1999).
\bibitem{H00} For a review, see, e.g., 
H. Hinrichsen. Adv. Phys. {\bf 49}, 815 (2000).
\bibitem{RPV00} M. Rossi, R. Pastor-Satorras, and A. Vespignani,
Phys. Rev. Lett. {\bf 85}, 1803 (2000).
\bibitem{KC1} J. Kockelkoren and H. Chat\'e, cond-mat/0306039.
\bibitem{PPun} S.-C. Park and H. Park (unpublished).
\bibitem{privateC} H. Chat\'e (private communication).
\bibitem{Hin010} H. Hinrichsen, Phys. Rev. E {\bf 63}, 036102 (2001);
Physica A {\bf 291}, 275 (2001); K. Park and I.-M. Kim, 
Phys. Rev. E {\bf 66}, 027106 (2002).
\bibitem{KC03} J. Kockelkoren and H. Chat\'e, Phys. Rev. Lett. {\bf 90},
125701 (2003).
\bibitem{Noh04} J. D. Noh and H. Park, Phys. Rev. E {\bf 69}, 016122 (2004).
\bibitem{Hin030} H. Hinrichsen, Physica A {\bf 320}, 249 (2003). 
\bibitem{BC03} G. T. Barkema and E. Carlon, 
Phys. Rev. E {\bf 68}, 036113 (2003).
\bibitem{HH04} M. Henkel and H. Hinrichsen, J. Phys. A {\bf 37}, R117 (2004).
\bibitem{PP04} S.-C. Park and H. Park, cond-mat/0406606.
\bibitem{JvWOT04} H.-K. Janssen, F. van Wijland, O. Deloubriere, and
U. C. T\"auber, Phys. Rev. E {\bf 70}, 056114 (2004).
\bibitem{RGall} M. Doi, J. Phys. A {\bf 9}, 1465 (1976); {\bf 9}, 
1479 (1976); P. Grassberger and M. Scheunert, 
Fortschr. Phys. {\bf 28}, 547 (1980); L. Peliti, J. Phys. (Paris) 
{\bf 46}, 1469 (1985); B. P. Lee, J. Phys. A {\bf 27}, 2633 (1994);
J. L. Cardy and U. C. T\"auber,  J. Stat. Phys. {\bf 90}, 1 (1998).
\bibitem{Aukrust} T. Aukrust, D. A. Browne, and I. Webman, 
Phys. Rev. A {\bf 41}, 5294 (1990).
\bibitem{DCA} I. Dornic, H. Chat\'e, and M. A. Mu\~noz, cond-mat/0404105.
\bibitem{bAK} D. ben-Avraham and J. K\"ohler, 
Phys. Rev. A {\bf 45}, 8358 (1992).
\bibitem{CAM} M. Suzuki and M. Katori, J. Phys. Soc. Jpn. {\bf 55}, 1 (1986);
M. Suzuki, {\it ibid}. {\bf 55}, 4205 (1987); 
J. Stat. Phys. {\bf 49}, 977 (1987).
\bibitem{CAMapp} H. Park, J. K\"ohler, I.-M. Kim, D. ben-Avraham, and S. Redner,
J. Phys. A {\bf 26}, 2071 (1993); A. L. C. Ferreira and S. K. Mendiratta, 
{\it ibid}. {\bf 26}, L145 (1993); 
N. Inui, Phys. Lett. A {\bf 184}, 79 (1993). 
\bibitem{CAMsol} N. Menyh\'ard and G. \'Odor, J. Phys. A {\bf 28}, 4505 (1995).
\bibitem{Szol2}  A. Szolnoki, cond-mat/0408114.
\bibitem{J93} I. Jensen, Phys. Rev. Lett. {\bf 70}, 1465 (1993).
\bibitem{PCP} R. Dickman, W. R. M. Rab\^elo, and G. \'Odor, 
Phys. Rev. E {\bf 65}, 016118 (2001); A. Szolonoki, {\it ibid}. {\bf 66},
057102 (2002).
\bibitem{mathematica} The conventional numerical integration methods allow us 
to find approximate solutions for larger $n$ \cite{Szol2}, 
but the accuracy is much more limited
($\sim10^{-10}$) within a reasonable computing time.
\bibitem{HS88} R. Bulirsch and J. Stoer, Numer. Math. {\bf 6}, 413 (1964);
M. Henkel and G. Sch\"utz, J. Phys. A {\bf 21}, 2617 (1988).
\bibitem{Szol3} The value of $\beta_2$  was independently reported by Szolnoki 
\cite{Szol2} very recently in the CMF scheme with the CAM analysis. 
\bibitem{2dPCPD} G. \'Odor, M. C.  Marques, and M. A. Santos, Phys. Rev. E 
{\bf 65}, 056113 (2002); S.-C. Park and H. Park (unpublished).
\end{thebibliography}
\end{document}